# Broadband telecom single-photon emissions from InAs/InP quantum dots grown by MOVPE droplet epitaxy

Shichen Zhang(张世晨)[1], Li Liu(刘丽)[1], Kai Guo(郭凯)[2], Xingli Mu(沐兴黎)[2,3], Yuanfei Gao(高远飞)[1], Junqi Liu(刘俊岐)[2,3], Fengqi Liu(刘峰奇)[2,3], Quanyong Lu(陆全勇)[1, *], Zhiliang Yuan(袁之良)[1]

[1] Beijing Academy of Quantum Information Sciences, Beijing 100193, China

[2] Laboratory of Solid-State Optoelectronics Information Technology, Institute of Semiconductors, Chinese Academy of Sciences, Beijing, 100083, China

[3] Center of Materials Science and Optoelectronics Engineering, University of Chinese Academy of Sciences, Beijing 100049, China

*Supported by National Natural Science Foundation of China (grant nos. 12494604, 12393834, 12393831, 62274014, 62235016 and 62335015) and National Key R&D Program of China (2024YFA1208900).

** These authors contributed equally: Shichen Zhang(张世晨), Li Liu(刘丽)

***Corresponding author. Email: luqy@baqis.ac.cn

The development of quantum materials for single-photon emission is crucial for the advancement of quantum information technology. Although significant advancement has been witnessed in recent years for single photon sources in near infrared band (λ~700-1000 nm), several challenges have yet to be addressed for ideal single photon emission at the telecommunication band. In this study, we present a droplet-epitaxy strategy for O-band to C-band single-photon source based semiconductor quantum dots (QDs) using metal-organic vapor-phase epitaxy (MOVPE). Via investigating the growth conditions of the epitaxial process, we have successfully synthesized InAs/InP QDs with narrow emission lines spanning a broad spectral range of λ~1200-1600 nm. The morphological and optical properties of the samples were characterized using atomic force microscopy and micro-photoluminescence spectroscopy. The recorded single-photon purity of a plain QD structure reaches ($g^{(2)}(0) = 0.16$), with a radiative recombination lifetime as short as 1.5 ns. This work provides a crucial platform for future research on integrated microcavity enhancement techniques and coupled QDs with other quantum photonics in the telecom bands, offering significant prospects for quantum network applications.

---

The deterministic emission characteristics of single photons, entangled photon pairs, and indistinguishable photons are important for efficient quantum light sources in applications such as quantum communication network and quantum computing[1-3]. Semiconductor epitaxial nanostructures have the most potential in the preparation of high purity single photon sources, and III-V compound semiconductor quantum dots (QDs) are considered to be almost perfect quantum emitters for the development and application of quantum networks[3-6].

To enable high-speed, long-distance quantum information transmission, QD-based light sources must operate with high brightness at telecom wavelengths, specifically in the O-band (1310 nm) or C-band (1550 nm), where optical fibers exhibit minimal transmission losses (0.35 dB/km and 0.2 dB/km, respectively)[7]. Furthermore, the relatively low background solar irradiance and Rayleigh scattering in the telecom C-band render QD light sources particularly suitable for free-space and satellite-based quantum communication systems[8, 9].

In recent years, the characteristics of single-photon sources with emission wavelengths below 1 μm based on QDs have reached near-ideal level[10-14], demonstrating significant advances in brightness, single-photon purity, indistinguishability, and entanglement fidelity[7, 12, 15-19]. Notably, recent studies utilizing tunable open microcavities coupled deterministically with high-quantum efficiency single QD, combined with tailored laser pulse excitation, have achieved a multi-photon error rate as low as 0.0205(6), photon indistinguishability reaching 0.9856(13), and a system efficiency of up to 0.712(18). This performance marks the first instance of surpassing the efficiency threshold required for scalable photonic quantum computing. This corresponds to a multi-photon suppression ratio exceeding 98%, signifying high-fidelity single-photon generation[14].

However, the development of semiconductor single-photon sources in the telecom band still lags significantly behind and requires further advancements to achieve comparable performance levels. While InAs/GaAs QDs with high photon emission purity have been successfully grew using the traditional Stranski-Krastanov (SK) mode[7, 20-22], a complex strain epitaxy process has to be performed to redshift the wavelength owing to the large lattice mismatching of 7.2% between InAs and GaAs[23]. This involves a growth of a metamorphic InGaAs buffer layer to adjust the degree of strain via changing lattice constant and material composition, thereby facilitating the redshift of the emission wavelength[24]. The strain dependence is closely linked to the pronounced fine-structure splitting (FSS), arising from the asymmetry in QD shapes, as well as the relatively high dot density (typically exceeding $10^9$ $cm^{-2}$)[25, 26], which together pose a major challenge for the prevailing S-K growth mode in realizing high-performance single-photon QD light sources.

In the InAs/InP system, due to the relatively lower lattice mismatch of 3.2%, the optical emission wavelength could be inherently tailored to the O-band to C-band ranges. However, it still facing the issue of high-density dot distribution using the S-K growth model. Droplet epitaxy (DE) recently has emerged as a highly promising alternative to the conventional SK growth mode. This process initially requires the regulation of substrate temperature and Group-III atomic flux deposition to form self-assembled droplets with specifically defined density and dimensions on the substrate surface. Subsequently, exposure to Group-V flux induces the direct crystallization of the droplets into III-V semiconductor QDs. The formation of QDs is not governed by the strain, thereby providing greater flexibility in the choice of epitaxial structures and materials[27, 28]. Another advantage is that near perfect symmetry has been achieved for the DE-grown QDs, which are ideal for generation of entangled photon pairs. In recent years, this technique has gained significant attention, particularly for the fabrication of low-density, high-symmetry QDs tailored for InAs/InP-based structures.

In this paper, we report high-quality low-density InAs/InP QDs for single-photon sources operating across a broad telecommunications band using quantum dot droplet epitaxy based on metal-organic vapor-phase epitaxy (MOVPE). The density and size of QDs are critical parameters that govern their optical and electrical characteristics. A low-density QDs distribution of ~$10^5$ $cm^{-2}$ on the sample alongside precise control of their size and shape is demonstrated, which enables the desired band-edge emission properties. This study investigates the control of low-density QD distribution and the optimization of their emission wavelengths to align with the communication band. The surface morphology and optical properties of the epitaxial QDs are characterized through atomic force microscopy (AFM) and photoluminescence (PL) spectroscopy. To verify the single-photon emission of the QD transitions, second-order intensity autocorrelation measurements under continuous wave excitation were performed and a low $g^{(2)}$ of 0.16 is obtained for the bare sample without using any optical cavities. The demonstrated high-quality InAs/InP QDs and their single photon emission property pave the way to the high-performance telecom single-photon sources when optical cavity is employed.

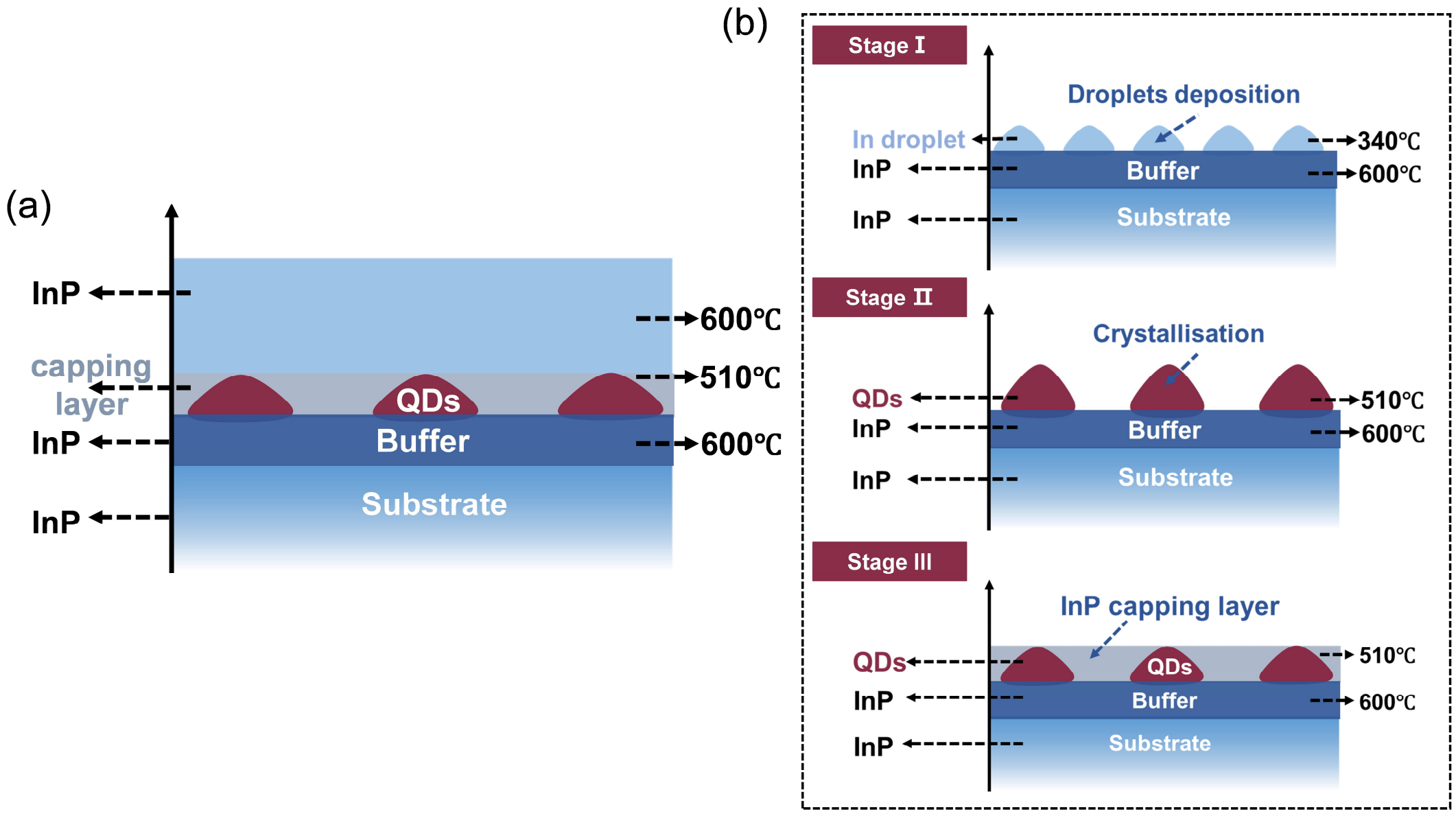

Fig. 1. (a) Schematic of the investigated QD structure. (b) Schematic of the droplet and QD formation sequence before growing InP cap layer (the top and middle panels), where the stage I represents the formation of In droplets and the stage II represents the crystallization of QDs under the arsine flow. The bottom panel displays the QDs buried by a capping layer.

The semiconductor QD samples were prepared using a DE technique in a MOVPE system equipped hydrogen ($H_2$) as the carrier gas. The precursors include trimethylindium (TMIn), arsine ($AsH_3$), and phosphine ($PH_3$) for the formation of In droplets, the crystallization of the droplets into QDs and the growth of InP cladding and cap layers, respectively. The epitaxial process starts with a 500 nm thick InP buffer layer growth at 600°C on a high-quality optical cavity-grown preferred surface InP (100) substrate, as illustrated in the growth sequence in Fig. 1.

The DE process consists of two main stages: the formation of In metal droplets (stage I) under a constant TMIn flow in the first place followed by their subsequent crystallization (stage II) under the exposure to an $AsH_3$ flow[27, 28]. In the droplet formation stage, the synergistic regulation of deposition temperature and growth interruption duration serves as a critical factor for controlling the density and uniformity of QDs. At lower temperatures, due to the limited migration length of adsorbed atoms, the formed In droplets have the characteristics of high density, small size and disordered spatial distribution. With the increase of temperature, the thermal activation energy will overcome the diffusion barrier, and the migration length of atoms will increase significantly, triggering the Ostwald ripening of the droplets. At this time, small droplets dissolve and transport to large droplets, so that the average diameter of the droplets expands and the density decreases.

However, when the substrate temperature was raised further, the excessive migration capability of atoms caused droplet coalescence and partial melting. Upon subsequent crystallization with As incorporation, these large droplets formed in this case further expanded, leading to the formation of two-dimensional microcrystalline structures in certain surface areas, thereby compromising the morphological uniformity of QDs[29]. After systematic optimization, we chose 340 °C as the deposition temperature for In and controlled the TMIn flow rate at 2.76 μmol/min for 35 s. Following droplet formation, we introduced a 25 s growth interruption before supplying arsenic (As) for crystallization. This interruption step promotes sufficient diffusion and effective attachment of In atoms to existing stable sites. This is beneficial to larger droplet formation with reduced density and narrowed droplet size distribution.

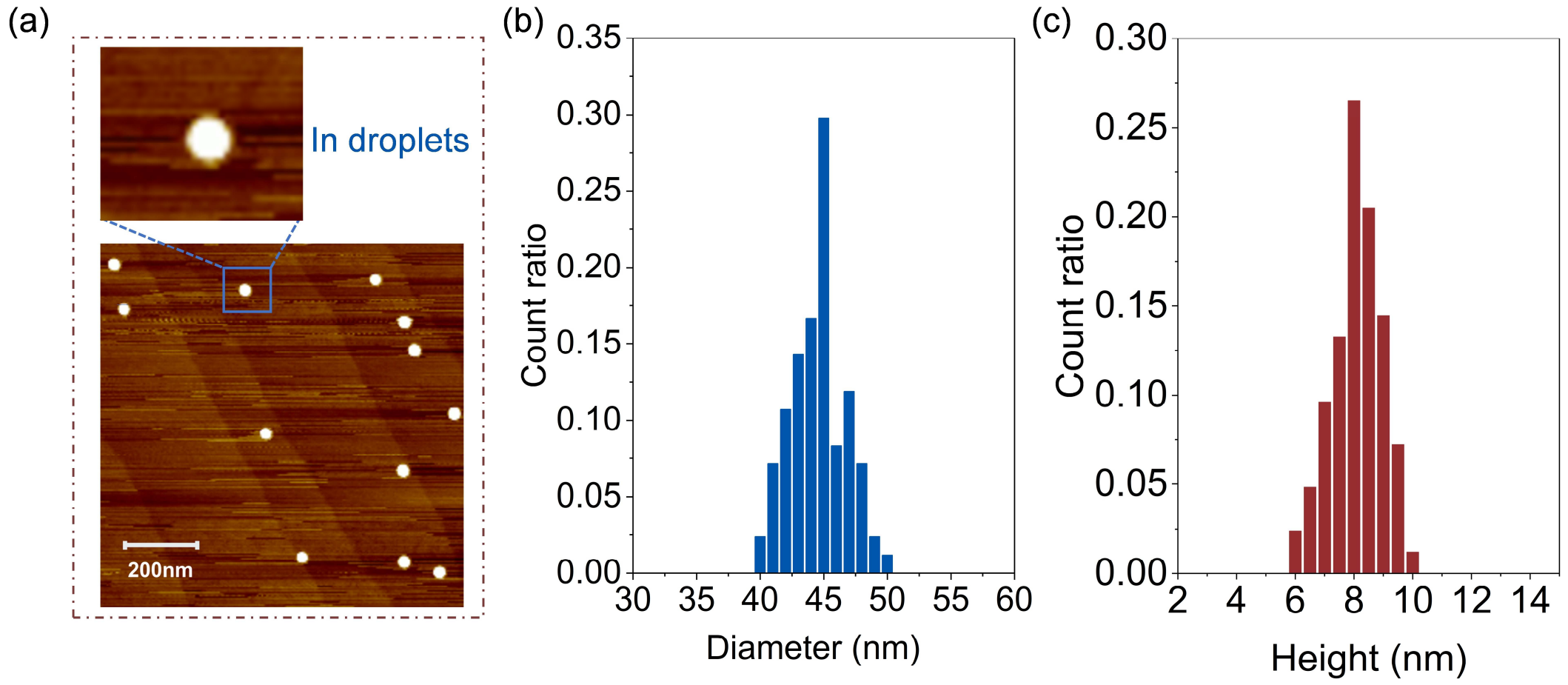

Fig. 2 (a) AFM images of the indium droplets on a bare InP substrate. Inset: zoomed-in AFM image of an indium droplet. (b) and (c) Distributions of diameters and heights of the indium droplets, respectively.

The morphology of the indium droplets under the aforementioned epitaxial parameters was systematically characterized using Atomic Force Microscopy (AFM), as shown in Fig. 2. The droplets exhibit a uniform distribution in both size and height under these growth conditions. To further analyze the In droplets, we obtained an indium droplet density of approximately $3\times10^8$ cm$^{-2}$ using multi-region statistical sampling (covering 5, 10 μm$^2$ regions). Fig. 2(b)-(c) show a histogram of the statistical distribution of size and height of 146 droplets. The results reveal that the droplets exhibit a highly concentrated size and height distribution without noticeable multi-peak or discrete features. The average diameter of the droplets is (45±5) nm and the average height is (8±2) nm. This distribution of size and height indicates that the chosen growth parameters, especially the growth interruption time, are reasonable and appropriate.

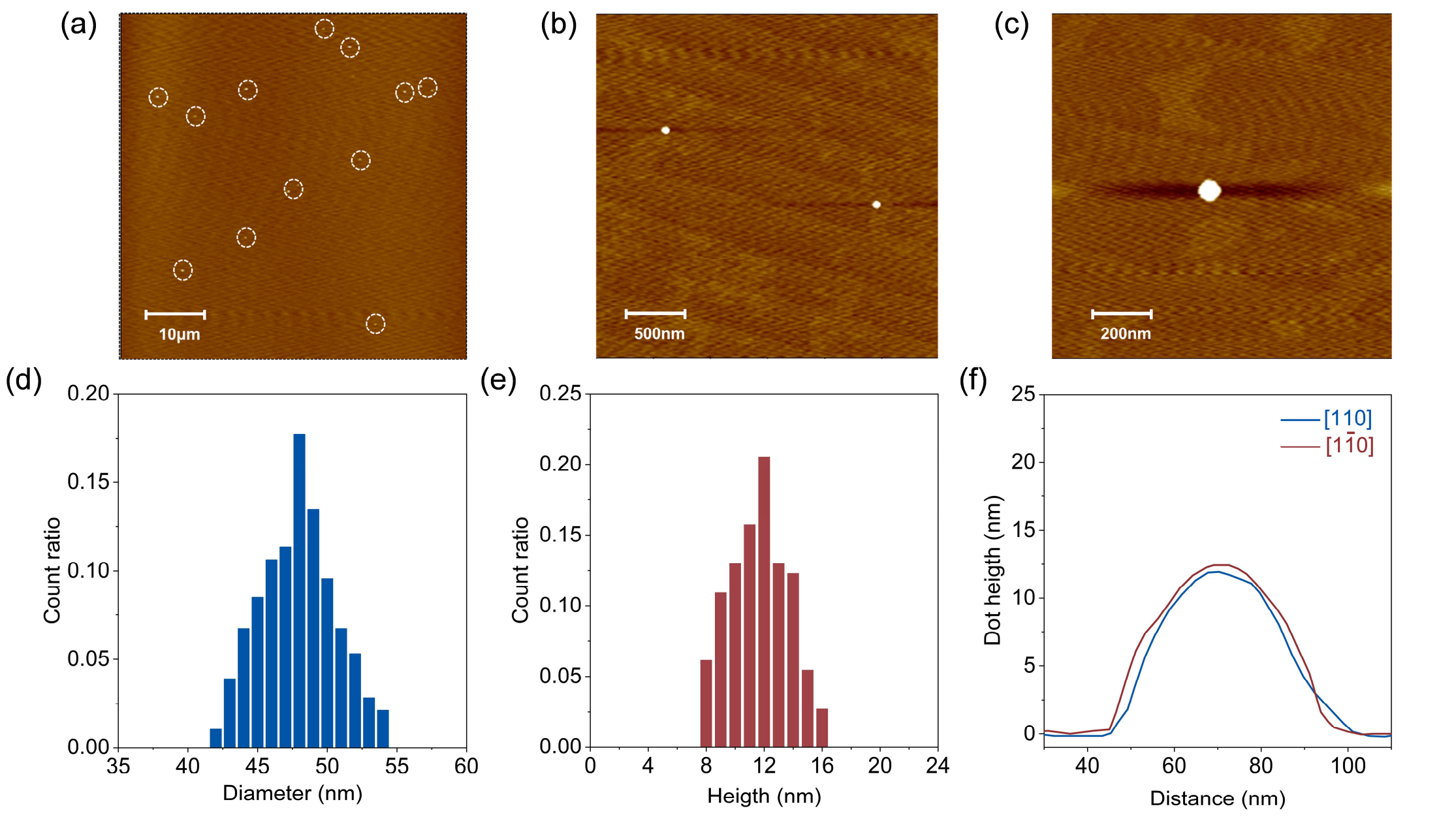

Fig. 3 AFM images of the QDs on a bare InP substrate: sample scanning range (a) 60×60 μm$^2$, (b) 3×3 μm$^2$ and (c) 1×1 μm$^2$. (d) and (e) Distributions of diameters and heights of QDs on bare InP, respectively. (f) Profile extractions of the QD shown in (c) along [110] and [1$\bar{1}$0] directions.

After formation of In droplets and a 25 s growth interruption, an $AsH_3$ flux of 26 μmol/min is applied for the crystallization of the droplets into InAs QDs which is the second key stage of DE (Fig. 1). The substrate temperature is then raised to 510 °C and maintained for 5 min to facilitate the QD crystallization. Fig. 3(a)~(c) shows AFM images of as-prepared free-standing InAs/InP QDs on the InP surface at different scan scales. Three main features can be clearly observed: 1) Low density distribution: Following a statistical approach analogous to that employed for In droplet density quantification, the areal density of QDs was systematically characterized across multiple regions of the wafer. The average QD density was determined to be approximately $5\times10^7$ cm$^{-2}$. Notably, as shown in Fig. 3(a), certain localized areas—within a scanning region of 60 × 60 μm$^2$—exhibit a significantly reduced QD density as low as $4\times10^5$ cm$^{-2}$. 2) Uniform size: The size of QD is roughly consistent across the scanned area, as verified using the built-in size measurement system of the AFM. This uniformity is particularly pronounced in the 3×3 μm$^2$ scan range shown in Fig. 3(b). 3) High symmetry of QD: The aspect ratio of the QDs was statistically analyzed through elliptical measurements using AFM. The average in-plane aspect ratio was found to be 1.06, with a standard deviation of 0.0487. These results demonstrate that the QDs exhibit high symmetry, as evidenced by the data obtained under the proposed growth methodology (Fig. 3). This high symmetry is essential for achieving an ideal single-photon source. Through the analysis of a large number of crystallized QDs, the typical diameter and height of the QDs at 510°C were (48±6) nm, and (12±4) nm, respectively. It is noteworthy that our profilometry results indicate no significant localized etching near the QDs, which is attributed to the chosen crystallization temperature of 510 °C. If a higher temperature had been selected, the InP surface would have become largely unstable due to the lack of phosphine supply and the formation of an In-rich surface configuration following the deposition of In droplets, leading to be dissolved[29, 30]. Employing buried QD epitaxial structure has been shown as an important factor affecting the shape and size of QDs, thereby their emission characteristics[31-36]. In this work, a thin InP capping layer was deposited on the samples prepared under the above-mentioned epitaxial conditions (at the crystallization temperature of 510 °C) to encapsulate the QDs. To achieve the target emission wavelength, capping layer thicknesses of 5, 10, 15, and 20 nm were tested. Subsequently, the growth temperature was ramped up to 600°C, and an 80 nm thick InP layer was deposited as the topmost layer (Fig. 1, the far-right panel). The growth of this layer at a higher temperature is beneficial to complete the structure of the QDs and provide good optical properties for the surrounding layer.

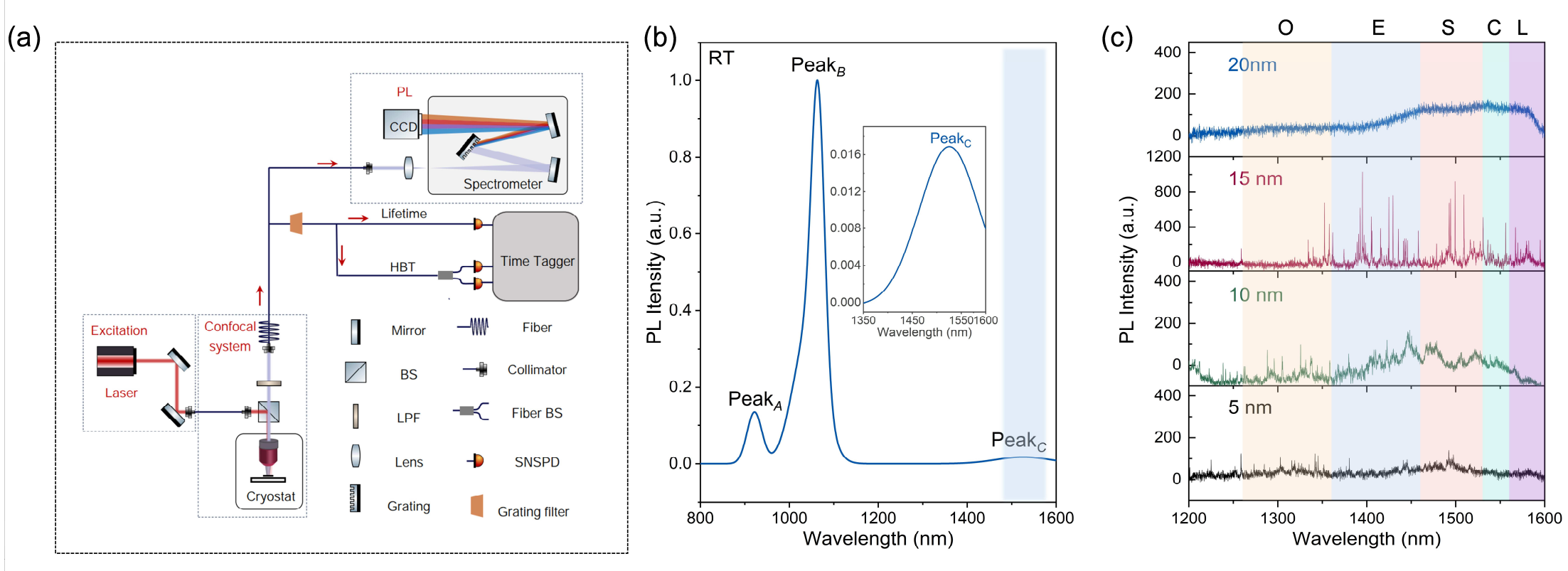

Fig. 4 (a) Schematic diagram of the custom-built optical performance characterization system for precise measurement and analysis of sample optical properties. (b) PL spectra of the QD sample with a 15 nm capping layer at room temperature (Inset: The vertical axis is compressed to highlight Peak$_C$). (c) μ-PL spectra of QD ensembles with varying capping layer thicknesses ranging from 5 to 20 nm under CW excitation (633 nm) at 4 K.

A custom-built confocal microscopic spectroscopy system was employed in this study to characterize the optical properties of quantum dot samples, as depicted in Fig. 4(a). The system consists of three main components: an optional pump laser source, a flow-type liquid helium cryostat equipped with an objective lens, and a multi-channel data acquisition system. The pump light is directed through a mirror and a beam splitter, then focused onto the sample surface in the cryogenic environment via the objective lens for excitation of the quantum dot emission. The excited fluorescence signal is collected by the same objective lens and coupled into different detection channels through the confocal optical path. These channels allow for spectral analysis, fluorescence lifetime measurement, and second-order autocorrelation function ($g^{(2)}(\tau)$) test, providing comprehensive data about the emission wavelength, radiative dynamics, and single-photon emission characteristics of the QDs.

All samples (the cap thickness of samples ranges from 5 nm to 20 nm) exhibit similar macroscopic PL line characteristics at room temperature (RT) under the high-power density excitation of a 633 nm diode laser, as shown in Fig. 4(b) using a 15 nm thin overburden sample as a representative case. The peak$_A$ (921.8 nm) is attributed to the InP substrate[29, 37], while the peak at 1064.3 nm (peak$_B$) is ascribed to the non-stoichiometric 2D layer formed on the surface during the temperature ramp and arsenic flux exposure, resulting from the arsenic-phosphorus exchange reaction. It is referred to as a "2D quasi-wetting layer" (WL) and is likely made of $InAs_xP_{1-x}$[29, 38]. In addition, a weak envelope is observed in the range above 1500 nm. Due to the relatively small lattice mismatch (~3%), the InAs/InP system naturally favors the formation of larger QDs. Therefore, we speculate that, as observed in previous studies, the peak at over 1500 nm (peak$_C$) originates from the contribution of low-density distributed QDs.

To further investigate whether the Peak$_C$ observed at RT arises from QD contributions with telecommunication band emission characteristics, the sample was placed in a flow-type liquid helium cryostat, maintained at a temperature of 4 K, for micro-photoluminescence (μ-PL) spectroscopy measurements. A continuous-wave (CW) laser (633 nm) was used to excite samples, with the excitation light focused to a spot size of 1.2 μm using an objective lens (NA = 0.65). Fig. 4(c) shows the spectra of the samples in the spectral of 1200-1600 nm range. For the samples with capping layer thickness of 5 nm and 10 nm, the emission lines are almost all distributed in the short-wavelength O and E communication bands. The number of individual emission lines that can be detected in the S-band and its red edge is small and the intensity is low, as shown in Fig. 4(c). As the thickness of the capping layer increases further upwards (15 nm), long wavelength emissions in the S, C, and even L bands are observed. When the thickness is increased to 20 nm, no valid single spectral line is detected even with further excitation intensity. This is attributed to the fact that as the thickness is further increased, the produced QDs are larger in size resulting a significant red shift, which is beyond the effective detection range of the spectrometer (cut-off to 1600 nm). Clearly, the cap layer thickness plays an important role in tuning the emitting wavelength of the QDs, a broadband emitting spectrum spanning the entire O, E, S, C and L bands with narrow lines spanning 1250-1600 nm is recorded for a cap layer thickness of 15 nm.

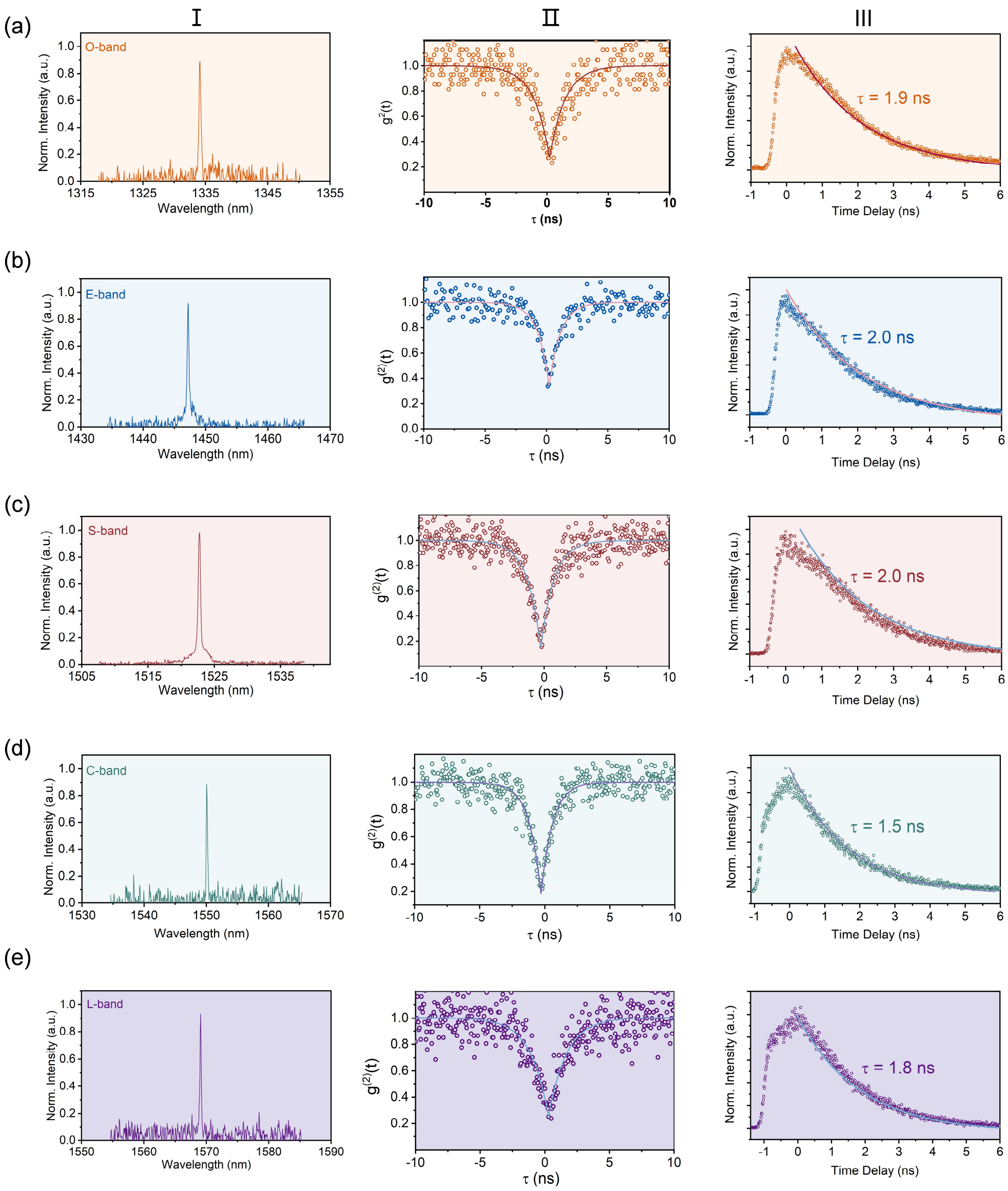

Fig. 5 Optical characterizations of the representative QDs from a 15-nm thick capping layer sample in different spectral ranges: (a) O-band, (b) E-band, (c) S-band, (d) C-band and (e) L-band, respectively. Panel I denotes the μ-PL spectra of each QD. The corresponding second-order correlation function measurements are shown in panel II (where circles represent the measured values and the solid line is the fit to the data). Panel III presents the fluorescence lifetime measurements of the QDs under 780 nm pulsed excitation.

To verify the single-photon emission characteristics of the QDs, second-order intensity autocorrelation measurements were performed on the single emission lines in each spectral band under continuous excitation with a 633 nm laser source. The second-order autocorrelation function values at zero delay under continuous-wave excitation conditions, for the O-band, S-band, E-band, C-band and L-band single-photon sources, are recorded in Fig. 5, respectively. The corresponding $g^{(2)}(0)$ through fitting are 0.28, 0.33, 0.16, 0.18, and 0.24, respectively, indicating pronounced single photon emissions spanning the entire O-L telecom wavelengths from the planar, cavity-free InAs QD structures. The single-photon performance would be further enhanced by certain nanocavities like the nanowires or elliptical Bragg grating (EBG), via the Purcell enhancement effect that intrinsically incorporating low-mode-volume optical confinement elements[10, 18, 39]. The PL lifetime measurements of QDs from corresponding spectral bands were also conducted under the 780 nm pulsed excitation. PL lifetimes of the QDs in each spectral band ranging from 1.5 to 2 ns are obtained by fitting the measured experimental data using an exponential decay model. The weak electron confinement in the InP/InAs material system and low local density of states of the photonic field lead to slow recombination rates compared to other material systems (such as GaAs/InAs) with cavity enhancements.[18, 40-43]. It is anticipated that, with the integration of subsequent microcavity designs (such as elliptical Bragg gratings and topological bulk cavity), the lifetime can be compressed to the order of hundreds of picoseconds.[39, 43]

In conclusion, we have prepared InAs/InP QDs for broadband telecom band single photon emissions by DE based on MOVPE. The characterization of the surface mophology by AFM shows that the density distribution of quantum dots as low as ~$4 \times 10^5$ $cm^{-2}$ can be achieved by optimizing the epitaxial temperature and other parameters. Covered by a 15-nm thin InP cap layer, the grown QD samples emits effective single-line emissions in a wide range of spectra of the entire O-L bands. The $g^{(2)}(0)$ of these QDs is as low as 0.16, indicating their excellent single-photon emission characteristics. The prepared InAs QDs embedded in InP claddings exhibit stable optical performance under continuous-wave excitation through multiple experimental runs, confirming the excellent fatigue resistance and process reproducibility. The MOVPE epitaxial technology offers inherent scalability for its mass production. The emission in the telecom band enables direct compatibility with existing fiber optic infrastructure and hybrid quantum-classical network architectures. These characteristics demonstrate that the demonstrated quantum dot light material system meets both stability and system integration requirements for quantum network applications, providing a reliable material platform for advancing quantum information technologies. By further integrating microcavity enhancement techniques or coupling the QDs to other optical devices, there is a promising and reliable pathway to rapid advancement of the quantum information technologies and their applications, particularly in the fields of quantum networking and secure communication.